\documentclass[reprint,amsmath,amssymb,aps,prb]{revtex4-1}
\pdfoutput=1
\usepackage{hyperref}
\usepackage{graphicx}
\usepackage[space]{grffile}
\usepackage{latexsym}
\usepackage{textcomp}
\usepackage{longtable}
\usepackage{multirow,booktabs}
\usepackage{amsfonts,amsmath,amssymb}
\usepackage{natbib}
\usepackage{url}
\hypersetup{colorlinks=false,pdfborder={0 0 0}}

\usepackage[utf8]{inputenc}
\usepackage[english]{babel}

\renewcommand {\vec} [1] {\mathbf{#1}}

\begin{document}

\title{Non-linear conductivity of metals from real-time quantum simulations}

\author{Xavier Andrade}
\author{S\'ebastien Hamel}
\author{Alfredo A. Correa}
\affiliation{Lawrence Livermore National Laboratory, 7000 East Avenue, Livermore, CA 94550, USA}

\begin{abstract}
We simulate bulk materials under strong currents by following in real-time the dynamics of the electrons under an electric field.
By changing the intensity of the electric field, our method can model, for the first time, non-linear effects in the conductivity from first principles.
To illustrate our approach, we show calculations that predict that liquid aluminum exhibits negative-differential conductivity for current densities of the order of \(10^{12}-10^{13}~\mathrm{A/cm^2}\).
We find that the change in the non-linear conductivity emerges from a competition between the accumulation of charge around the nuclei that increases the scattering of the conduction electrons, and a decreasing scattering cross-section at high currents.%
\end{abstract}%

\maketitle



In 1827, Georg Ohm established the law that carries his name by discovering that when a voltage is applied over a piece of a certain material, it will carry a current proportional to that voltage. 
The proportionality constant depends on the geometry of the conductor, and a property of the material: the conductivity.
Since then, the conductivity has become one of the most basic quantities that characterize a material. In fact, we usually classify solids into three mayor groups: metals, semiconductors and insulators, based on their ability to conduct electricity.

As it is common with 19th century ``laws'', Ohm's law is more a first order approximation to a more complex behavior than a fundamental property: many materials do not follow Ohm's law, diodes for example, and for those that do, we can expect that for large enough voltages the conductivity of a material will change.
Probably the simplest case of non-linear conduction is the dielectric breakdown, which occurs when an insulator becomes conductive if it is exposed to a very strong electric field.
In some materials it can even happen that the current decreases as the voltage is increased, leading to the phenomenon of negative differential conductivity (also known as negative differential resistance)~\cite{Ridley_1963,Volkov_1969,Pamplin_1970}. 
Negative differential conductivity has been observed mainly in semiconductors, like GaAs or GaN, and is used for technological applications~\cite{Gunn_1963,Gru_inskis_1999}. 
It has also been reported in molecular electronic devices~\cite{Chen_1999,Xue_1999,Dalgleish_2006,Perrin_2014} and nano-structures~\cite{Lyo_1989,Rinkio__2010,Zheng_2010,Wu_2012,Du_2012,Lin_2015}.
As far as we know, negative differential conductivity has not been observed in metals due to the high fields that would be required~\cite{Pamplin_1970}.

For some materials the conductivity is difficult to measure, for example in the case of matter under extreme conditions that are not easy to replicate in an experimental setup.
In situations like this, the computational study and prediction of transport properties is an essential tool to understand experiments, and to help in the development of technological applications like thermoelectric power generation~\cite{Chen_2003} or molecular electronics~\cite{L_rtscher_2013}.

The theoretical determination and prediction of linear transport coefficients in materials has been studied from many different points of view~\cite{Drude_1900,Ziman_1961,ashcroft1976solid,Mahan_1990}.
In most cases, the electrons are the ones that carry electricity, so the conductivity is determined by the electronic structure of material. 
This means that an accurate theoretical description of conduction requires a quantum mechanical treatment of the electrons.
From a first-principles perspective, the electrical conductivity of crystalline systems is usually calculated using perturbation theory, by combining density-functional theory and the Kubo-Greenwood formula~\cite{Kubo_1957,Greenwood_1958}.
This method only yields the conductivity in the low current regime, so up to now non-linear effects in the conductivity have been out of reach for first principles methods.

In this work, we present a new method to obtain the conductivity of bulk materials that not only allows us to access the standard conductivity, even in the limit of zero frequency, but also the effective non-linear conductivities for high-intensity currents. 
To do it, we apply an electric field, and we simulate the response of the material by following the quantum dynamics of the electrons in real-time.
By controlling the intensity of the applied field we gain access to the non-linear regime.
We can also monitor the time- and space-resolved electronic and current densities to try to understand the microscopic nature of conduction and how non-linear effects appear.


We apply our method to the calculation of the conductivity of aluminum at high temperatures and under strong electric fields. 
Under these conditions, we find that the conductivity depends on the intensity of the current, leading to negative differential conductivity. 
To validate our results, we develop a simple model based on ion-in-jellium scattering theory that shows a similar dependence of the conductivity with the current density. 
We also find that the inhomogeneous accumulation of electronic charges in the system shows distinct patterns for the different conduction regimes, allowing us to gain insight into the cause of the changes in conductivity.

\section{Theory}

To describe the dynamics of the electrons we use an effective Schr\"odinger equation in the form of time-dependent density functional theory (TDDFT)~\cite{Runge_1984} by solving the time-dependent Kohn-Sham (KS) equation (Gaussian atomic units, where \(\hbar=m_e=e=1\), are used throughout)
\begin{equation}\label{eq:ks}
\mathrm{i}\frac{\partial}{\partial t}\varphi_{i}(\vec{r},t) = \hat{H}[n](t)\,\varphi_{i}(\vec{r},t)
\end{equation}
where \(\varphi(\vec{r},t)\) are the time-dependent KS orbitals, \(\hat{H}\) is the KS single-particle Hamiltonian that reproduces the electron-electron interaction, and \(n\) is the electronic density.

Periodic boundary conditions are the preferred way to deal with condensed-matter systems. 
Usually the electric field enters the Schr\"odinger equations through an external potential field in the Hamiltonian; however, the scalar potential associated to a uniform electric field does not satisfy the periodic boundary conditions. 
Instead, we use a gauge where the uniform electric field is generated using a uniform, but time-dependent, vector potential as~\cite{Bertsch_2000}
\begin{equation}
\vec{E}(t) = -\frac{1}{c}\frac{\partial \vec{A}(t)}{\partial t}\ .
\end{equation}
%
%
%
This additional vector potential is coupled to the electrons through the kinetic energy in the KS Hamiltonian
\begin{equation}
\hat{H}[n](t) = \frac12\left(\vec{\nabla} - \frac1c\vec{A}(t)\right)^2 + v_{\mathrm{ext}}(\vec{r}, t) + v_{\mathrm{hxc}}[n](\vec{r})\ ,
\end{equation}
where $v_\text{ext}$ is the potential generated by ions and \(v_{\mathrm{hxc}}[n]\) includes the Hartree, exchange and correlation potentials (HXC), that replicate the electron-electron interaction in the KS picture.


The fundamental observable in our approach is the time-dependent macroscopic current density \(\vec{J}(t)\). 
In a single-particle picture, the gauge-invariant current density can be calculated as
\begin{equation}
\label{eq:current}
\vec{J}(t) = -\frac{\mathrm{i}}{\Omega}\int \mathrm{d}\vec{r}\,\sum_{j}^N\varphi_{i}(\vec{r},t)\left[\hat{H}(t), \hat{\vec{r}}\right]\varphi_{i}(\vec{r},t)\ .
\end{equation}
where \(\Omega\) is the volume of the unit cell. 
This generalized form based on the commutator has the advantage that it is explicitly invariant under translations, 
and that it is correct when \(v_\text{ext}\) contains non-local pseudopotential terms.

The basic macroscopic quantity for characterizing the capacity of a medium to conduct electricity is the conductivity, \(\sigma\). It relates, to first order, a monochromatic electric field, \(\vec{E}(\omega)\), of frequency \(\omega\) with the current density it generates at that frequency, \(\vec{J}(\omega)\), by Ohm's law ~\cite{LANDAU_1984}
\begin{equation}
\label{eq:ohm}
\vec{J}(\omega) = \sigma(\omega)\vec{E}(\omega)\ ,
\end{equation}
If we consider Ohm's law in real-time~\cite{Allen_2006}, we can obtain the conductivity from an electron dynamics simulation by applying a time-dependent electric field \(\vec{E}(t)\), and measuring for each time the induced current \(\vec{J}(t)\). 
It is convenient to perturb the system using a so-called \emph{kick}~\cite{Yabana_1996}, an electric field of the form 
\(\vec{E}(t) = \vec{E}_0\delta(t)\), where \(\vec{E}_0\) determines the intensity and direction of the field. 
By inserting the kick electric field into Ohm's law, we find that the conductivity can be calculated as
\begin{equation}
\sigma(\omega) = \frac{1}{E_0}\int_0^\infty\mathrm{d}t\,e^{-\mathrm{i}\omega t} J(t)\ .
\label{eq:conductivity}
\end{equation}
Here we assume that the conductivity is a scalar, as the systems we are interested in this work are isotropic on average.
For a more general case where the conductivity is a tensor, three different simulations are required, with \(\vec{E_0}\) pointing in different directions.

We implemented this scheme in the real-space TDDFT code Octopus~\cite{Castro_2006,Andrade_2015}. 
The computational cost of our implementation scales quadratically with the number of atoms and does not require unoccupied states, while linear-response approaches typically scale cubically and require the calculation of unoccupied states. 
This fact, combined with the high parallel efficiency of real-time TDDFT~\cite{Andrade_2012,Draeger_2016}, allows us to simulate supercells with hundreds of atoms with a reasonable computational cost. 
For the real-time TDDFT simulations in this article we use the adiabatic local density approximation (ALDA), a grid spacing of $0.425~\mathrm{a.u.}$, and a time step of $0.1~\mathrm{a.u.} = 0.00242~\mathrm{fs}$.

\begin{figure}[h!]
\begin{center}
\includegraphics[width=1\columnwidth]{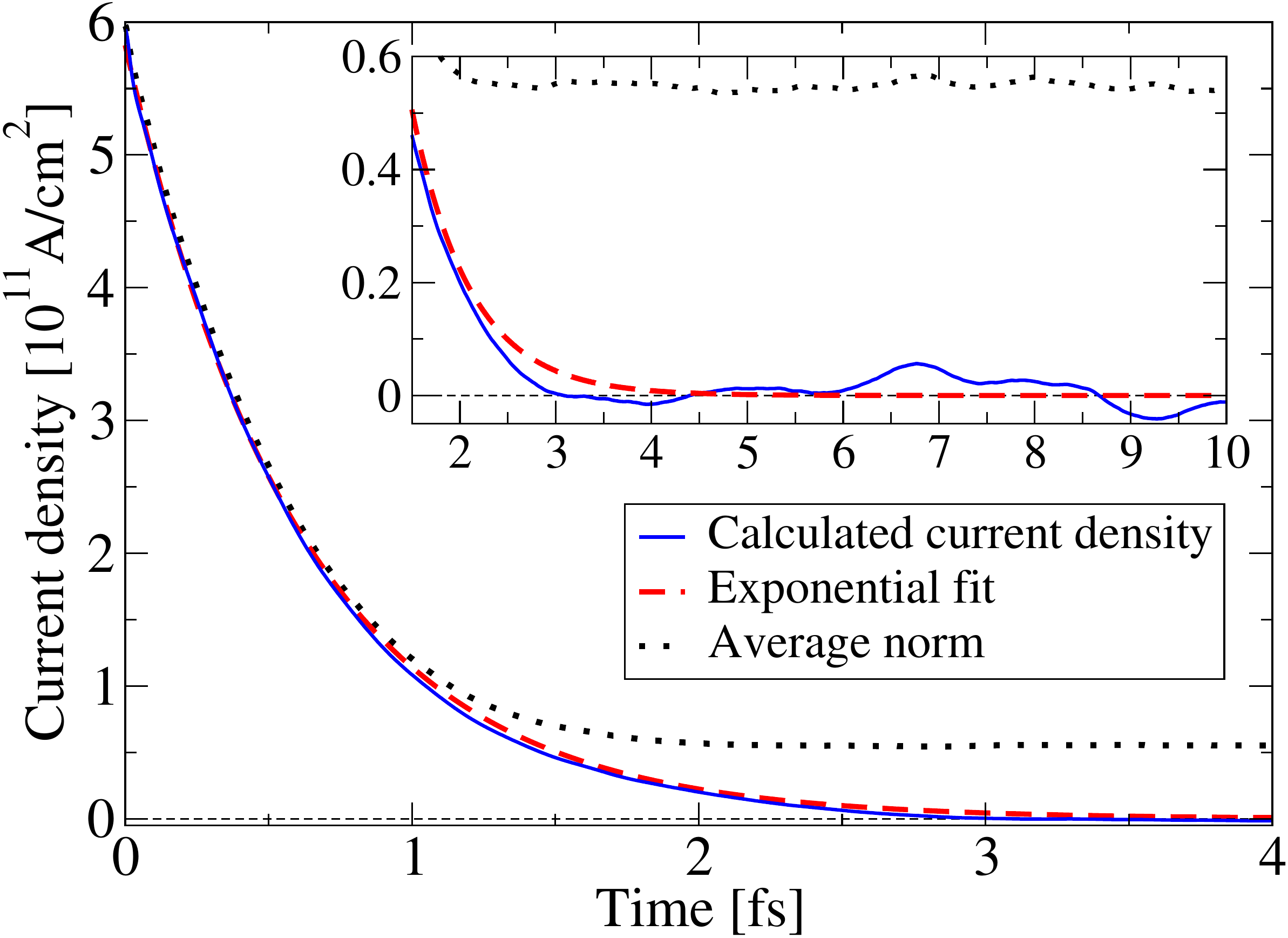}
\caption{{\label{fig:current_one} 
Current density vs. time after a kick of $0.1~\mathrm{a.u.}$ (continous blue line) for a specific ionic configuration of a supercell with 256 Aluminum atoms in the liquid phase. 
The inset shows in detail the current fluctuations around zero for long times.
The red segment line shows an exponential fit to the current, based on the range $0<t<0.5~\mathrm{fs}$.
The dotted black line shows the spatial average of the absolute value of the local current density, which illustrates how the macroscopic current density decays due to a randomization of the microscopic local current density, which retains a finite intensity.%
}}
\end{center}
\end{figure}

\begin{figure}[h!]
\begin{center}
\includegraphics[width=1\columnwidth]{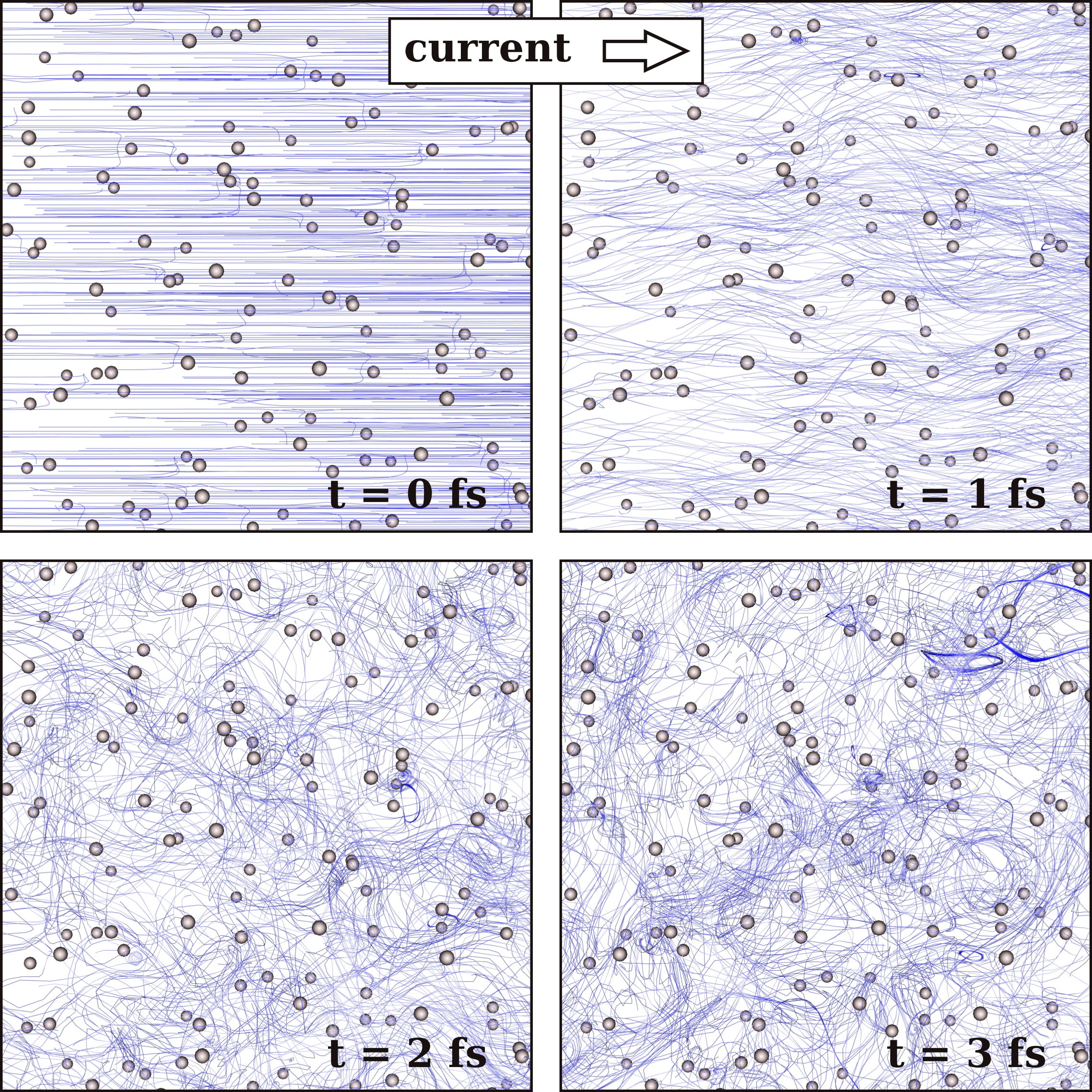}
\caption{{\label{fig:streamlines} Stream lines of the current density for different times for a kick of \(0.1~\mathrm{a.u.}\) in liquid aluminum. 
Initially the lines are parallel and follow the direction of the current (from left to right), as time passes the current density becomes more disordered and isotropic. 
This disorder in the direction of the local current density is partially responsible for making the average current density vanish.
The stream lines represent the trajectories of fictitious particles under the current density as a velocity field.%
}}
\end{center}
\end{figure}

\section{Results}

As a test case for our formalism we consider liquid aluminum at high temperature (\(0.5~\text{eV} \sim 5802~\text{K}\)) at solid density (\(2.7~\mathrm{g}/\mathrm{cm}^3\)). 
The liquid state is chosen to simplify the problem as in this state, the intrinsic disorder avoids pathologies that can arise in a perfectly crystalline system and its associated translation symmetries.
Our simulation supercell contains 256 atoms and we use the gamma point for Brillouin zone sampling.
To account for the ionic temperature, we perform a Born-Oppenheimer molecular dynamics (MD) simulation at the target temperature and take snapshots of the ionic positions.


We start by showing the results for a single simulation using one snapshot of the MD run.
In Fig.~\ref{fig:current_one} we plot the current density evolving in time after a kick of $0.1~\mathrm{a.u.}$ 
As seen in the figure, the current steadily decays towards zero. 
Still, small fluctuations of the current are observed after it has decayed (as seen on the inset of Fig.~\ref{fig:current_one}).
If we fit an exponential to the initial part of the current density ($t < 0.5~\mathrm{fs}$), we can see that the current is very close to an exponential function.

\begin{figure}[h!]
\begin{center}
\includegraphics[width=1\columnwidth]{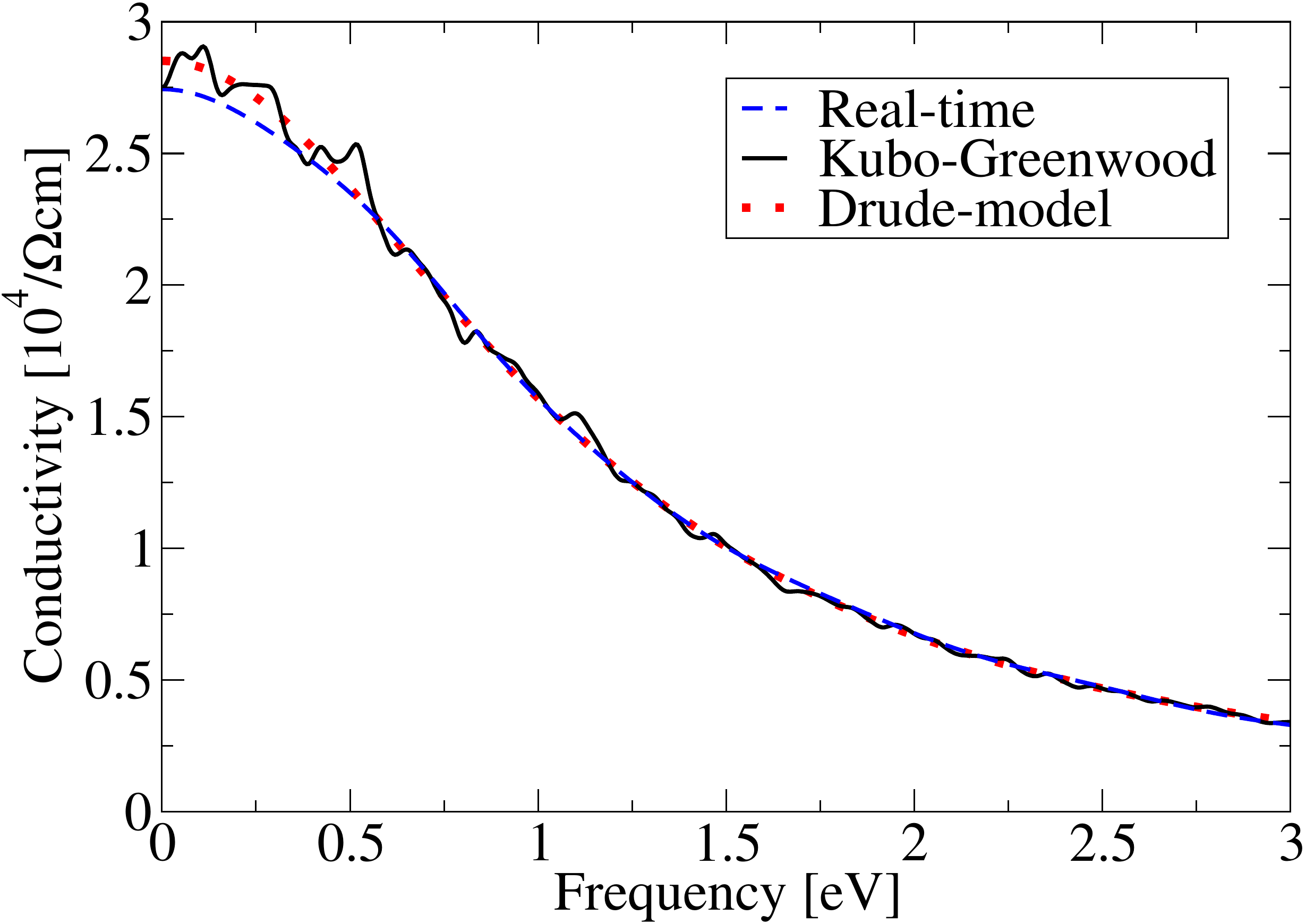}
\caption{{\label{fig:conductivity_one} Frequency-dependent conductivity in the low frequency regime  (blue segmented line), calculated as a Fourier transform of the time-dependent current density (Eq.~\ref{eq:conductivity}). 
We compare with the frequency-dependent conductivity calculated by the Kubo-Greenwood method (black continuous line), for the same ionic snapshot. 3000 bands are used, which allows for vertical excitation energies of up to 40 eV. 
Additionally, a Drude fit, obtained from the Fourier transform of the exponential fit of the real time current, is presented for comparison (red dotted line).
These results are for a single snapshot for a fair comparison between the two methods.%
}}
\end{center}
\end{figure}

\begin{figure}[h!]
\begin{center}
\includegraphics[width=1\columnwidth]{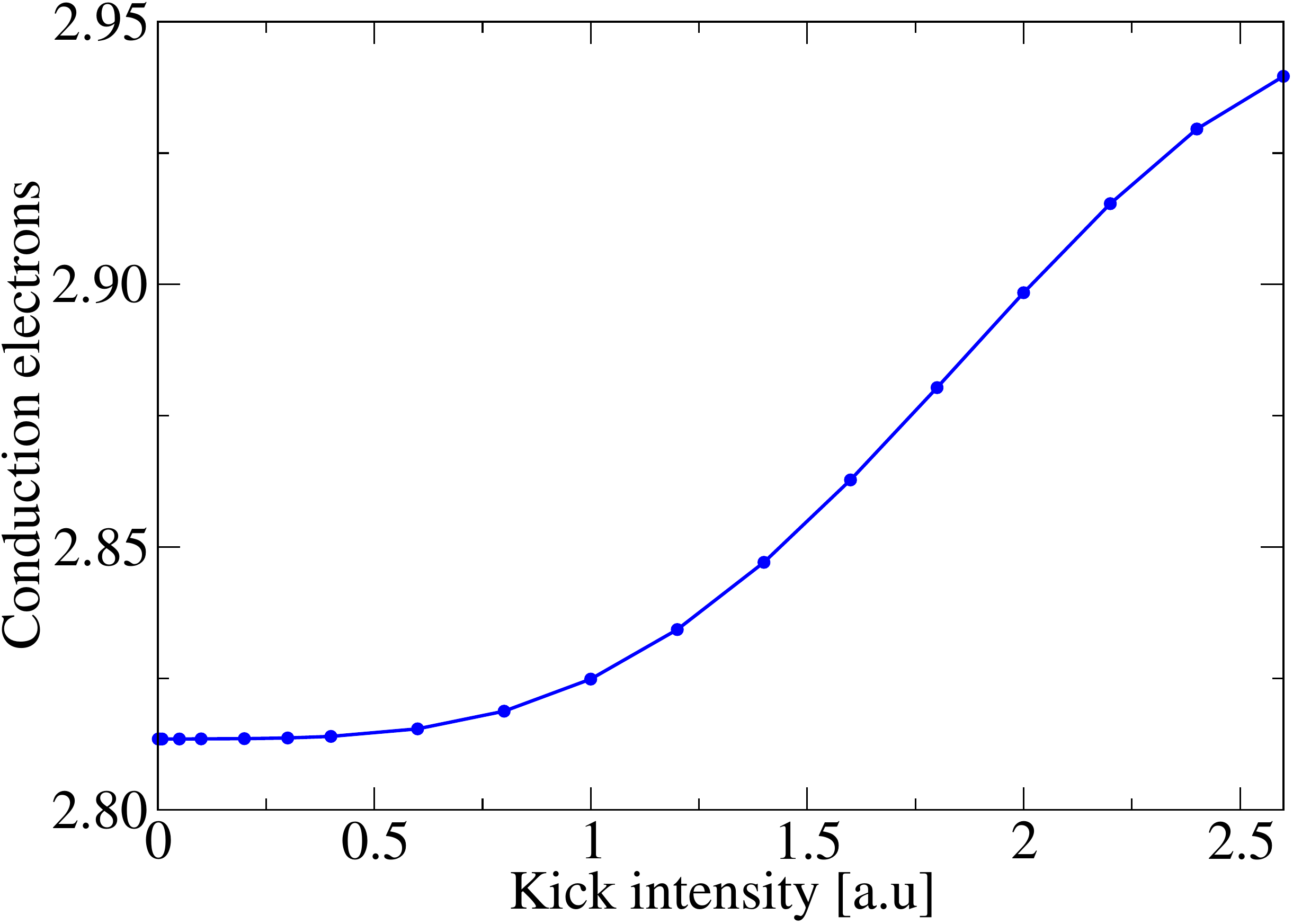}
\caption{{\label{fig:initial} Ratio of the initial current per atom with respect to the applied kick. This number is equivalent to the fraction of the electrons per atom that participate in the current at time zero. The value grows as the intensity of the kick increases and approaches 3, the valence electrons in aluminum.%
}}
\end{center}
\end{figure}

The real-time simulation allows us to get some insight into the nature of the decay process. 
Since the current density is the spatial average of the microscopic current density, the current density might decay because the microscopic current density is decaying to zero at each point, which is to be expected of a final time-reversible state. 
Alternatively, it is possible that the microscopic current density becomes disordered, making the average zero while retaining a finite value at each point.

To differentiate between those two scenarios we consider the microscopic current density (defined as the integrand in Eq.~\ref{eq:current}).
In Fig.~\ref{fig:current_one} we include the average value of the norm of the microscopic current density, and in Fig.~\ref{fig:streamlines} we plot the streamlines of the current density for different times.

Initially, both the current density and its average norm decay in a similar fashion. 
However, after some time the average norm remains constant while the average current continues decaying.
This tells us that both mechanisms are at play: initially (\(t < 1~\mathrm{fs}\)) there is a local decay of the magnitude of the current, but later  (\(t > 1~\mathrm{fs}\)) the disorder and spatial randomness of the current is responsible for nullifying the average current. 
The immediate conclusion is that the final state, although is time-reversible on its macroscopic averages is still a time-dependent quantum state without time-reversal symmetry. 
We note, however, that given the conservative and time-reversible nature of the equations (Eq.~\ref{eq:ks}) and the accuracy of our propagator~\cite{Castro_2004}, 
if we invert time and propagate the system backwards, starting from the final quantum state, we observe the same dynamics in reverse, with the spontaneous regeneration of the current.

To obtain the frequency-dependent conductivity we Fourier-transform the time-dependent current density, according to Eq.~\ref{eq:conductivity}. 
We can do a numerical Fourier transform of the current, or analytically transform the exponential fit, that gives us a Drude-like form for the conductivity. 
Both results are shown in Fig.~\ref{fig:conductivity_one}, where we compare them with the results from Kubo-Greenwood~\cite{Kubo_1957,Greenwood_1958} for the same single ionic configuration. 
It can be observed that there is a good agreement between real-time TDDFT and Kubo-Greenwood. 
In Kubo-Greenwood the conductivity is a sum over discrete transitions between orbitals, so a broadening of $0.03~\mathrm{eV}$ is applied to produce a continuous curve, still, the conductivity is much rougher than the real-time approach. 
The small differences in the results can be attributed to the additional approximations introduced in Kubo-Greenwood: the truncation of the sum over unoccupied states, and the lack of a full self-consistent response.

The results we have shown so far correspond to a single snapshot, that serve to illustrate the detailed properties of our new method. 
However, if we want to make predictions, we need to average the results over multiple ionic configurations obtained from MD.
We took 200 ionic snapshots from a $12~\mathrm{ps}$ equilibrated MD run using \textsc{VASP}~\cite{Kresse_1996} at the LDA level of approximation to the exchange-correlation functional and a 2x2x2 Monkhorst-Pack grid. 
We used a MD time step of 2 fs.   
For each sample we calculated the evolution of the current for different kick intensities from \(0.001\) to \(2.6~\mathrm{a.u.}\), that allows us to access linear and non-linear effects.

\begin{figure}[h!]
\begin{center}
\includegraphics[width=1\columnwidth]{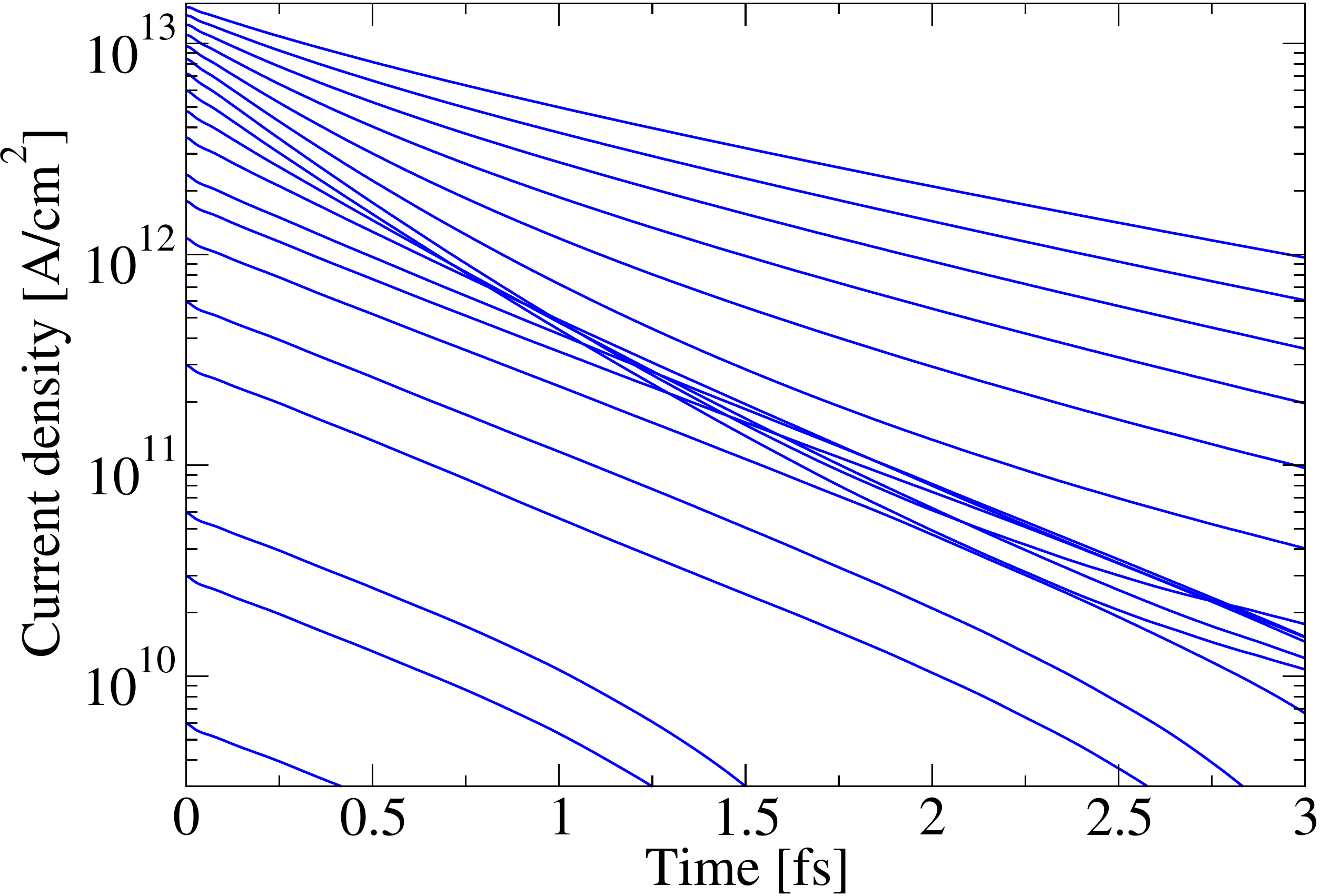}
\caption{{\label{fig:current_log} 
Time-evolution of the current density for a family of initial conditions with different intensities of electric \emph{kick} intensities, and therefore initial current densities, averaged over ionic configurations.
For initial current below $10^{12}~\mathrm{A/cm^2}$ we observe a decay with the same time constant, as expected from a linear response.
For larger perturbations we enter into a non-linear regime that is evidenced by the changing slopes of the graphs.%
}}
\end{center}
\end{figure}

We first check how the initial current depends on the kick by calculating the ratio of the initial current density over the kick intensity, which we plot in Fig.~\ref{fig:initial}.
Since the current corresponds to the momentum of the electrons, it can be shown that this ratio is essentially the amount of charge that participates in the conduction. 
In a disordered metallic system this value, when normalized per atom, should be close to the valence charge. We see that for small kicks, below $0.5~\mathrm{a.u.}$, a constant fraction of around \(2.81\) electrons per atom are excited, while for higher intensity kicks this value starts to grow and saturates towards 3 for the kick considered, which is the valence charge for aluminum.

We now consider the actual decay behavior to determine the linearity of the response with respect to the intensity of the perturbation.
For each kick value we obtain the current density, averaged over configurations, as a function of time. 
The values are shown, in logarithmic scale, in Fig.~\ref{fig:current_log}. 
If we consider that the slope of the current decay gives us an idea of the value of the conductivity, we can clearly see that there is a non-linear behavior. 
As the intensity of the initial current increases, the decay is less exponential-like and the rate of decay changes, first increasing and then decreasing.

\begin{figure}[h!]
\begin{center}
\includegraphics[width=1\columnwidth]{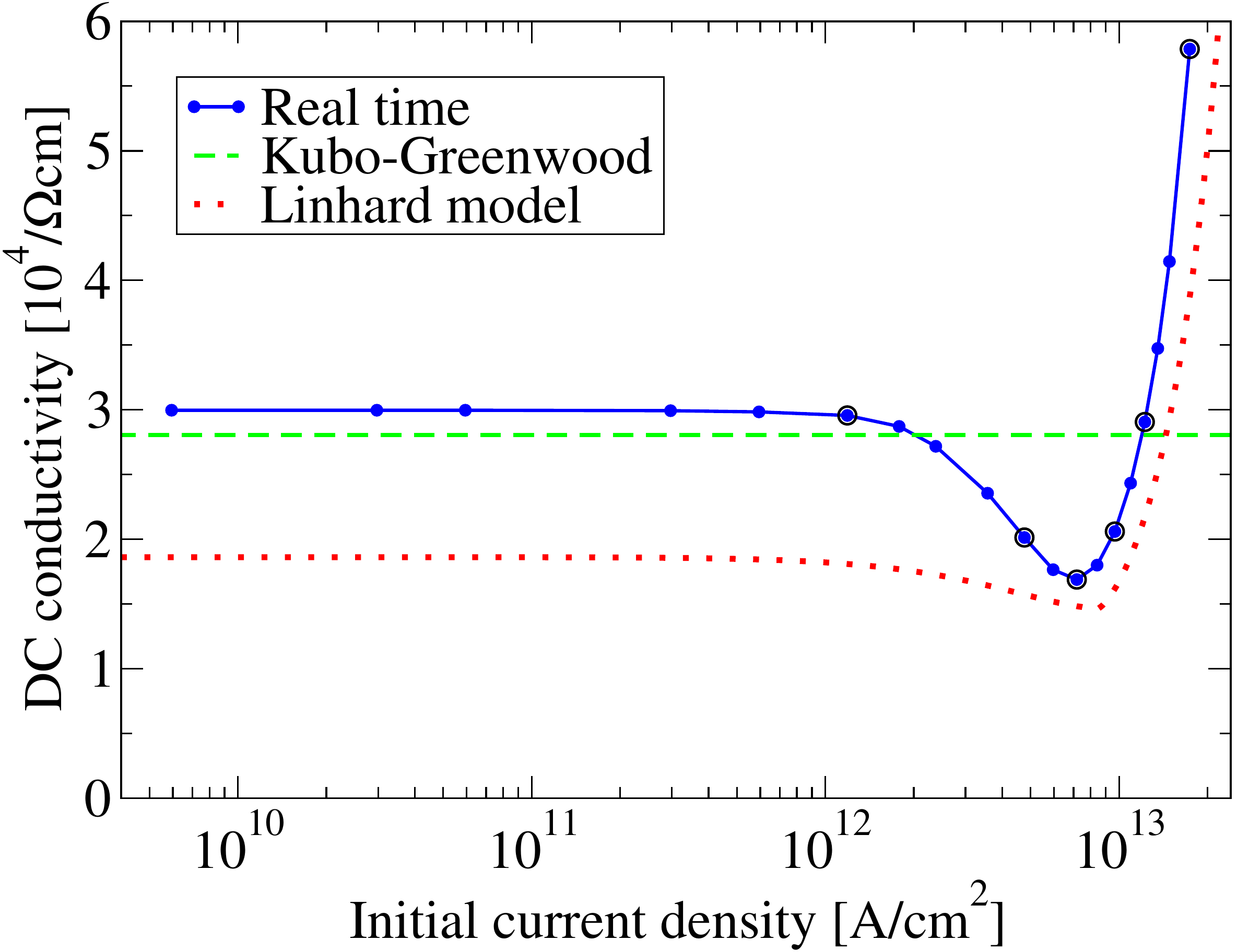}
\caption{{\label{fig:nonlinear} 
DC-conductivity as a function of the current density.
For low current the value of the conductivity agrees with Kubo-Greenwood and at large currents there is a minimum of conductivity at $\sim 7\times 10^{12}~\mathrm{A/cm^2}$.
At even larger currents the conductivity increases as the electrons become less affected by scattering processes.
The points encased in a black circle indicate the values of the current shown in Fig.~\ref{fig:blobs}.%
}}
\end{center}
\end{figure}

To further analyze the non-linearity, we determine the DC conductivity for each curve in Fig.~\ref{fig:current_log} by an exponential fit to the initial part of the time-dependent current (\(t<0.5\ \mathrm{fs}\)).
The resulting conductivity \emph{vs.} initial current is shown in Fig.~\ref{fig:nonlinear}. 
The conductivity exhibits a strong dependency with the current density.
For low currents, we obtain a conductivity \(3.0\times 10^4~(\mathrm{\Omega\,cm})^{-1}\) that is close to the $\omega\to 0$ extrapolated Kubo-Greenwood value of \(2.8\times 10^4~(\mathrm{\Omega\,cm})^{-1}\). 
These values also agree with previous theoretical results~\cite{Ashcroft_1965,Silvestrelli_1999,Desjarlais_2002,Recoules_2005,Vl_ek_2012}. 
For higher intensities, the conductivity exhibits a minimum value of \(\sim1.7\times 10^4~(\mathrm{\Omega\,cm})^{-1}\) for \(\sim7.2\times 10^{12}~\mathrm{A/cm^2}\) which corresponds to a kick, or electronic velocity, of \(\sim1.2~\mathrm{a.u.}\), comparable to the Fermi velocity we calculate for this system: \(0.92~\mathrm{a.u.}\) For larger currents the conductivity increases rapidly.


\begin{figure}[h!]
\begin{center}
\includegraphics[width=1\columnwidth]{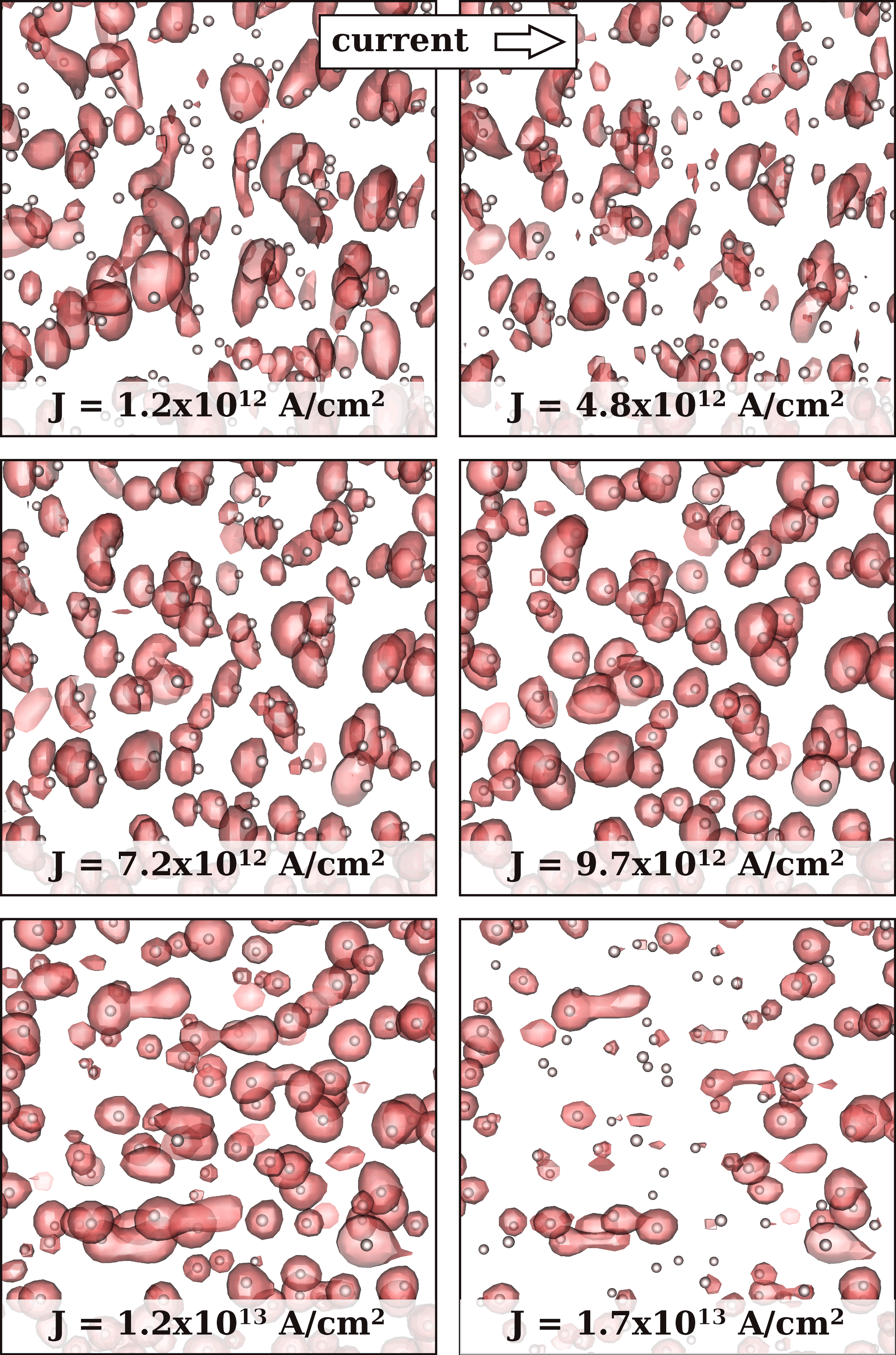}
\caption{{\label{fig:blobs} Regions of accumulation of electronic charge for different values of the initial current, for a single snapshot. The red surfaces represent isosurfaces of the difference of the density at time \(0.242~\mathrm{fs}\) with respect to the ground state density. The value of the isosurfaces are proportional to the perturbation intensity and are \(0.0025\), \(0.01\), \(0.015\), \(0.02\), \(0.025\) and \(0.03~\mathrm{a.u.}\), respectively.%
}}
\end{center}
\end{figure}

To explore the origin of the non-linearity in the conductivity, we plot in Fig.~\ref{fig:blobs} the change in the electronic density for different initial currents for a particular snapshot and a fixed time. It can be seen that there is a clear change in regime depending on the initial currents. For low currents there is some small accumulation of charge, mainly in the interatomic regions. For larger currents the charge starts to accumulate close to the nuclei. For \(\sim7.2\times 10^{12}~\mathrm{A/cm^2}\), which is close to the conductivity minimum, the accumulation of the charge occurs in front of the atoms. As the current is increased the charge accumulation regions are centered around the atoms, and in some cases some appendices form behind the atoms. Finally, for large currents these regions of charge accumulation start to disappear.

We theorize that the generation of these regions of charge accumulation increase the amount of scattering that the electrons are subject to, reducing the conductivity. At the same time, for larger currents we expect the conductivity to increase, as the electrons become ballistic and are less affected by scattering processes. The combination of these two effects explain the shape of the conductivity curve that we see in Fig.~\ref{fig:nonlinear}.

Based on the Linhard ion-in-jellium scattering theory \cite{Lindhard1954}, we have developed a simple model to approximate the conductivity of metals. 
The formula proposed is derived from electronic stopping theory~\cite{Schleife_2015}, where the collective force on the ions is reinterpreted as a deacceleration of the electronic stream. An adjustable spatial cutoff is introduced to capture a finite radius of action, \(R_\text{s}\), for each individual scatterer. The conductivity, as a function of the current, in our model is given by
\begin{equation}
\sigma(J) = \frac{\pi}2 \frac{J^2}{ Z n } \left(\int^\infty_{1/R_\text{s}} \frac{\mathrm{d}k}{k} \int^{kJ/n}_0\mathrm{d}\omega\, \omega\, \mathrm{Im}\left[ \frac{-1}{\varepsilon(\omega, k)} \right] \right)^{-1}
\end{equation}
where \(Z\) is the number of participating electrons per ion, and \(\varepsilon(\omega, k)\) is some appropriate model of the dielectric response. In the model, \(Z\) is assumed to be \(3\), based on the valence of aluminum and backed up by the results of Fig.~\ref{fig:initial}. \(R_\text{s}\) is set to one fourth of the average interatomic distance, which replicates the high current portion of the results.

The model is formally justified at large currents as it is perturbative in \(Z/J\); however as we can see from Fig.~\ref{fig:nonlinear}, it replicates qualitatively well the shape of the simulated conductivity, reproducing the main features of the curve: a constant value for low currents, a minimum close to the current corresponding to the Fermi velocity, and a conductivity that increases rapidly for large values. This divergence behaves as \(\sim J^2/\log J\), as one would expect from the Born~\cite{Born1926} and Coulomb-logarithm scattering theory~\cite{Sivukhin1966}.

It is remarkable that this duality exists between the theories of the perturbative and non-perturbative regimes: the non-linear conductivity in the high-current limit \(\sigma(J\to\infty)\) appears to  be related to the linear dielectric response \(\varepsilon(\omega, k)\). Of course, both theories are linear in different parameters, the former is perturbative in \(Z/J\) and the latter is perturbative in the external field.

\begin{figure}[h!]
\begin{center}
\includegraphics[width=1\columnwidth]{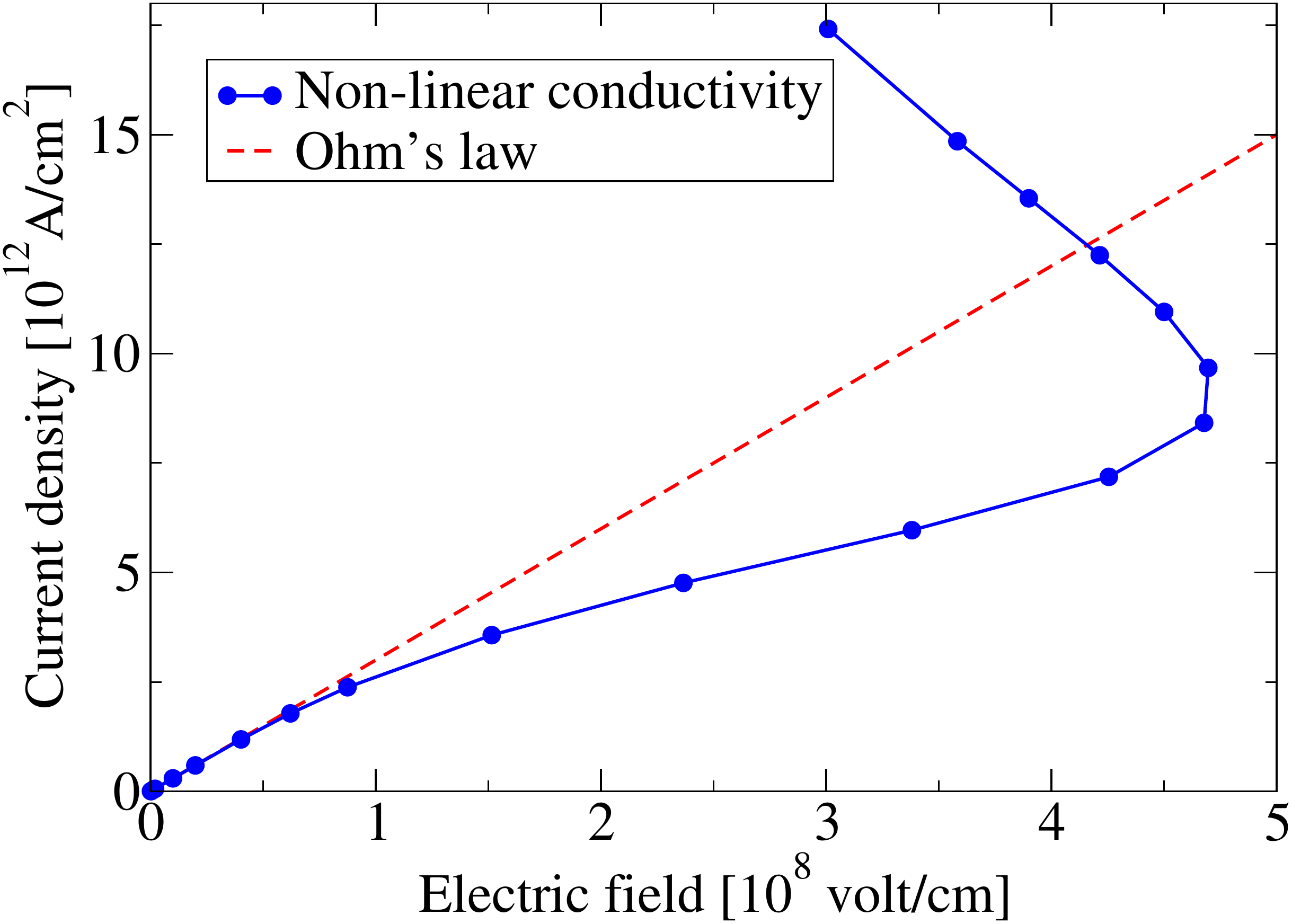}
\caption{{\label{fig:evsj} Current density vs electric field curve for liquid aluminum calculated from the current dependency of the conductivity. This curve predicts that aluminum is an S-shape negative-differential conductor for very high currents and fields.%
}}
\end{center}
\end{figure}

Based on the conductivity results for each current density, we can obtain the electric field \emph{vs.} current density curve shown in Fig.~\ref{fig:evsj}. It shows that our simulations predict a negative differential conductivity in liquid aluminum at \(5800~\mathrm{K}\) for electric fields of the order of \(10^8~\mathrm{V/cm}\). The induced current density in this regime is of the order of \(10^{12}-10^{13}\ \mathrm{A/cm^2}\). We have not found any indication that this effect has been directly measured, as such regimes do not seem to be experimentally reachable today. The negative differential conductivity that we find is of S-type, or current-controlled~\cite{Ridley_1963}. In practice, the multi-valued current density for a given field causes the system to separate in different spatial regions with different currents. We expect that the curve will shift again to the right, completing the \textit{S} shape, for higher currents, as semi-core electrons will become involved in the conduction.

\section{Conclusions} 

We have presented a new approach to study electrical conduction, and other current-related phenomena, in metallic materials. Rather than studying response properties, we directly apply an electric field on the system and follow, in real-time, the evolution of the current density. This method has practical advantages over the standard Kubo-Greenwood approach, traditionally used to calculate the electrical conductivity of materials. Interestingly, we show a fundamentally irreversible process in the simulation of a quantum system of many electrons from microscopically reversible equations. 

The most important feature of the new method, however, is that it uses finite electric fields and currents, allowing us to explore non-linear effects in the conductivity.
We apply this feature to the conductivity of liquid aluminum for very-high current densities (\(> 10^{12} \mathrm{A}/\mathrm{cm}^2\)). 
We find that in our simulations the conductivity of aluminum starts to change under these conditions, leading us to predict that aluminum presents negative differential conductivity. 
To the best of our knowledge, negative differential conductivity has not been observed in metals, probably due to the large electric fields and currents involved. 
However, such non-linear effects could be relevant in microscopic regions where the current density could reach much higher values than the macroscopic average. 
It is also possible that the non-linear regime becomes accessible when the material is under more extreme conditions of temperature or pressure; for example
an analogous effect, the `two-stream instability', is known to develop in classical plasmas~\cite{Chen_1984}. 
Future applications of our method will be to study the conductivity of semiconductor and insulator materials, where negative differential conductivity is well known, and routinely used in electronic devices~\cite{Gunn_1963,Gru_inskis_1999}.

Our real-time method is certainly not restricted to electrical conductivity in metals, which is one of the simplest applications of modeling currents in materials. 
This first application opens the path for studying other phenomena that up to now have been beyond the reach of realistic quantum simulations. For example, an interesting extension of our approach would be to study electronic thermal conductivity and mixed electrical-thermal transport properties, so that we can simulate, understand and predict non-linear effects in heat transport, including negative-differential heat conduction~\cite{Hu_2011,Zhu_2012,Zhou_2016}.
In our framework, it will require the introduction of recent developments of the description of thermal properties in TDDFT~\cite{Eich_2014a,Tatara_2015,2016arXiv160705464E}.
We could also study the molecular dynamics of a material under an electric field or a current. 
A particularly interesting application of this idea is electromigration~\cite{Ho_1989}: the movement of ions in a material due to the momentum transfer from the electronic current. 
Electromigration is very important from a technological point of view, as it is one of the main causes of failure in integrated circuits~\cite{Malone_1997}.
Our ultimate goal is to build a general theoretical and computational tool that can combine different types of perturbations and observables into time-resolved simulations.
By accessing combined responses and non-linear effects, this tool would give us an unprecedented capacity to study different phenomena that are of interest in condensed matter physics, material science, and plasma physics.

\begin{acknowledgments}

This work was performed under DOE Contract No. DE-AC52-07NA27344.
Work performed at the Energy Dissipation to Defect Evolution Center, an Energy Frontier Research Center funded by the U.S. Department of Energy (Award Number 2014ORNL1026). 
Computing support for this work came from the LLNL Institutional
Computing Grand Challenge program. LLNL IM release number
LLNL-JRNL-715217.
\end{acknowledgments}

\bibliographystyle{apsrev4-1}
\bibliography{bibliography/converted_to_latex.bib%
}

\end{document}